\documentclass[aip,amsmath,amssymb,reprint]{revtex4-2}

\usepackage{graphicx}
\usepackage{dcolumn}
\usepackage{bm}
\usepackage[utf8]{inputenc}
\usepackage[T1]{fontenc}
\usepackage{mathptmx}
\usepackage{booktabs}
\usepackage{color}
\usepackage{multirow}
\usepackage{notes2bib}

\begin{document}

\title{Time evolution of natural orbitals in \textit{ab initio} molecular dynamics}

\author{Alejandro Rivero Santamar\'{i}a}
\email{alejandro.rivero@univ-lille.fr}
\affiliation{Univ. Lille, CNRS, UMR 8523 – PhLAM – Physique des Lasers Atomes et Molécules, F-59000 Lille, France}

\author{Mario Piris}
\email{mario.piris@ehu.eus}
\affiliation{Donostia International Physics Center (DIPC), 20018 Donostia; Euskal Herriko Unibertsitatea (UPV/EHU), PK 1072, 20080 Donostia; Basque Foundation for Science (IKERBASQUE), 48009 Bilbao, Euskadi, Spain}

\date{\today}

\begin{abstract}
This work combines for the first time \textit{ab initio} molecular dynamics (AIMD) within the Born-Oppenheimer approximation, with a global natural orbital functional (GNOF), an approximate functional of the one-particle reduced density matrix. The most prominent feature of GNOF-AIMD is the ability to display the real-time evolution of natural orbitals, providing detailed information on the time-dependent electronic structure of complex systems and processes, including reactive collisions. The quartet ground-state reaction N($^4$S) + H$_2$($^1\Sigma$) $\rightarrow$ NH($^3\Sigma$) + H($^2$S) is taken as validation test. Collision energy influences on integral cross sections for different initial ro-vibrational states of H$_2$ and rotational-state distributions of NH product are discussed, showing a good agreement with previous high-quality theoretical results.
\end{abstract}

\maketitle

One-particle reduced density matrix (1RDM) functional theory\citep{Gilbert1975, Levy1979, Valone1980} is an alternative formalism to both density functional and wavefunction based methods. A pragmatic approach results in approximate functionals of the 1RDM in its diagonal form, that is, the use of natural orbitals (NOs) and its occupation numbers (ONs) as the fundamental variables, which define a natural orbital functional (NOF).\citep{Piris2007} An important issue is that the approximate NOFs continue to depend on the two-particle RDM (2RDM),\citep{Donnelly1979} so it is necessary to consider their functional N-representability.\citep{Ludena2013,Piris2018d} The NOF theory is currently an active research field,\citep{Sutter2023,DiSabatino2023,Cioslowski2023,Lew-Yee2022,DiSabatino2022,Senjean2022,Bousiadi2022,Wang2022,Schade2022,Rodriguez-Mayorga2022,Liebert2022,Yao2021,Gibney2021,Liebert2021,Schmidt2021,Schilling2021,Mercero2021,Quintero-Monsebaiz2021,Rodriguez-Mayorga2021,Cioslowski2020,Giesbertz2020,Benavides-Riveros2020,Mitxelena2020a,Mitxelena2020b} which can already be applied to large molecular systems of general chemical interest\citep{Lemke2022,Lew-Yee2023a} using open-source software like DoNOF.\citep{Piris2021a,Lew-Yee2021} Exhaustive reviews of NOF-based methods can be found elsewhere.\citep{Pernal2016, Schade2017, Mitxelena2019} 

Knowledge of the analytical derivatives with respect to nuclear coordinates in NOF theory\citep{Mitxelena2017,Mitxelena2020c} allows routine calculation of molecular structures and related properties. This ability to calculate gradients analytically also makes NOF-based \textit{ab initio} molecular dynamics (AIMD)\citep{marx_hutter_2009} simulations feasible. The basic idea is to calculate the forces acting on the nuclei from the calculations of the electronic structure as the molecular dynamics trajectory is generated. In this way we avoid determining in advance the potentials of interatomic interactions that can have serious drawbacks in chemically complex systems where the bond pattern changes qualitatively in the course of dynamics.

In many cases, the electrons respond almost instantaneously to nuclear motion. In such situations, the Born-Oppenheimer (BO) approximation allows us to decouple the electronic and nuclear problems, defining a potential energy surface (PES). In this work, we focus on the gas-phase dynamics of molecules in their electronic ground state (GS), where the BO approximation is generally accurate. There are other problems, for example, when the dynamics starts in an excited electronic state, where strong couplings between two or more PESs occur, requiring quantum treatment of both nuclei and electrons. Non-adiabatic molecular dynamics\citep{Tapavicza2013,Curchod2018} is the method of choice to model these processes, but that is outside the scope of this work.

Nowadays, the most reliable approaches to deal with electron correlation in AIMD are multireference methods, such as the complete active space self-consistent field (CASSCF)\citep{Roos1980,Roos2007} and its combination with second-order perturbation theory (CASPT2).\citep{Roos1982,Andersson1990,Andersson1992,Pulay2011} Unfortunately, multireference methods can suffer from high computational cost, sensitivity to active space selection, and instabilities in calculations due to changes in active space orbitals that can lead to energy discontinuities along the trajectory. Consequently, few simulations using CASPT2 nuclear gradients have been reported to perform non-adiabatic AIMD for reduced molecular models  \citep{Park2017,Luzon2019} and, so far, there are no examples of convergent CASPT2-AIMD calculations in reactive collisions.

NOFs offer an intermediate cost between multireference methods and common density functionals. In fact, approximate NOFs have been shown to be more accurate than their electron density-dependent counterparts and to have better scaling with respect to the number of basic functions than wavefunction-based methods for systems with a large amount of strong non-dynamic correlation. Interestingly, NOFs further corroborate the motivation behind the floating occupation molecular orbital complete active space configuration interaction (FOMO-CASCI) method, a promising alternative to CASSCF in dynamics simulations.\citep{Slavicek2010,Hollas2018}

The aim of this article is to present for the first time a BO AIMD based on an approximate NOF, that is, the nuclei will propagate according to the classical equations of motion, in an adiabatic PES obtained by solving in each time step the quantum mechanical electronic structure problem using a NOF. The price to pay is that the correlation lengths and relaxation times that are accessible are smaller than those available through standard molecular dynamics, but we can handle chemically complex systems, avoid the dimensionality bottleneck that arises when PESs are calculated in advance,\citep{marx_hutter_2009} and see the real-time evolution of NOs and their ONs during complex dynamics, obtaining detailed information about the time-dependent electronic structure of such processes.

As a functional, we will use the recently proposed global NOF (GNOF)\citep{Piris2021b} for electronic systems with any value of spin regardless of external potential. It has been demonstrated\citep{Mitxelena2022} that GNOF provides a good balance between static and dynamic electronic correlations leading to accurate total energies while preserving spin, even for systems with a highly multi-configurational character. In addition, GNOF has proven to be successful in dealing with aromaticity\citep{Mercero2023} and charge delocalization error.\citep{Lew-Yee2023b} GNOF correlates all electrons into all available orbitals for a given basis set; which today is not possible for large systems with current wavefunction-based methods. Regarding the dynamics, the NOs vary along the trajectory, adapting at each time step to the most favorable interactions of the corresponding nuclei configuration. In the following, we briefly describe GNOF.

We consider a mixed quantum state (multiplet) of an N-electron molecule with $\mathrm{N_{I}}$ spin-unpaired electrons, $\mathrm{N_{II}}=\mathrm{N-N_{I}}$ spin-paired, and total spin $S$. We focus on the state of highest multiplicity: $2S+1=\mathrm{N_{I}}+1$.\citep{Piris2019} For the whole ensemble, the expected value of $\hat{S}_{z}$ is zero; so the spin-restricted theory is adopted, i.e., a single set of orbitals is used for $\alpha$ and $\beta$ spins: $\varphi_{p}^{\alpha} \left(\mathbf{r}\right) = \varphi_{p}^{\beta} \left(\mathbf{r}\right) = \varphi_{p} \left(\mathbf{r}\right)$. Accordingly, all spatial orbitals are double occupied in the ensemble, so that occupancies for particles with $\alpha$ and $\beta$ spins are equal: $n_{p}^{\alpha}=n_{p}^{\beta}=n_{p}.$

The orbital space is divided in turn into two subspaces: $\Omega=\Omega_{\mathrm{I}}\oplus\Omega_{\mathrm{II}}$. $\Omega_{\mathrm{II}}$ is composed of $\mathrm{N_{II}}/2$ mutually disjoint subspaces $\Omega_{g}$, each of which contains one orbital $\left|g\right\rangle$ with $g\leq\mathrm{N_{II}}/2$, and $\mathrm{N}_{g}$ orbitals $\left|p\right\rangle $ with $p>\mathrm{N_{II}}/2$, namely,
\begin{equation} 
\Omega_{g}=\left\{ \left|g\right\rangle ,\left|p_{1}\right\rangle ,\left|p_{2}\right\rangle ,...,\left|p_{\mathrm{N}_{g}}\right\rangle \right\}, \quad \Omega_{g} \in \Omega_{\mathrm{II}}. \label{OmegaG}
\end{equation} 
Taking into account the spin, the total occupancy for a given subspace $\Omega_{g}$ is 2, which is reflected in the following sum rule:
\begin{equation}
\sum_{p\in\Omega_{g}}n_{p}=n_{g}+\sum_{i=1}^{\mathrm{N}_{g}}n_{p_{i}}=1,\quad g=1,2,...,\frac{\mathrm{N_{II}}}{2}.\label{sum1}
\end{equation} 
In general, $\mathrm{N}_{g}$ can be different for each subspace as long as it describes the electron pair well. For convenience, we usually take it the same for all subspaces $\Omega_{g}\in\Omega_{\mathrm{II}}$. The maximum possible value of $\mathrm{N}_{g}$ is determined by the number of basis functions ($\mathrm{N}_{B}$).
From (\ref{sum1}), it follows that
\begin{equation}
2\sum_{p\in\Omega_{\mathrm{II}}}n_{p}=2\sum_{g=1}^{\mathrm{N_{II}}/2}\left(n_{g}+\sum_{i=1}^{\mathrm{N}_{g}}n_{p_{i}}\right)=\mathrm{N_{II}}.\label{sumNpII}
\end{equation}
On the other hand, $\Omega_{\mathrm{I}}$ is composed of $\mathrm{N_{I}}$ mutually disjoint subspaces, each of which contains only one orbital $\left|g\right\rangle$. This spatial orbital is individually occupied, $n_{g}^{\alpha} + n_{g}^{\beta} = 2n_{g} = 1,$ but we do not know whether the electron has $\alpha$ or $\beta$ spin. This leads to
\begin{equation}
2\sum_{p\in\Omega_{\mathrm{I}}}n_{p}=2\sum_{g=\mathrm{N_{II}}/2+1}^{\mathrm{N_{\Omega}}}n_{g}=\mathrm{N_{I}}.\label{sumNpI}
\end{equation}
In Eq. (\ref{sumNpI}), $\mathrm{N}_{\Omega} = \mathrm{N_{II}}/2+\mathrm{N_{I}}$ denotes the total number of suspaces in $\Omega$. Note that the orbitals are arranged as follows: from 1 to $\mathrm{N_{II}/2}$, there are the strongly occupied orbitals belonging to $\Omega_{\mathrm{II}}$, followed by the single-occupied orbitals belonging to $\Omega_{\mathrm{I}}$ ranging from $\mathrm{N_{II}}/2+1$ to $\mathrm{N}_{\Omega}$. After that, there are the weakly occupied orbitals of each $\Omega_{g} \in \Omega_{\mathrm{II}}$. Taking into account Eqs. (\ref{sumNpII}) and (\ref{sumNpI}), the trace of the 1RDM is verified to be equal to the number of electrons: 
\begin{equation}
2\sum_{p\in\Omega}n_{p}=2\sum_{p\in\Omega_{\mathrm{II}}}n_{p}+2\sum_{p\in\Omega_{\mathrm{I}}}n_{p}=\mathrm{N_{II}}+\mathrm{N_{I}}=\mathrm{\mathrm{N}}.\label{norm}
\end{equation}
It is essential to note that orbitals undergo changes throughout the optimization process to find the most favorable orbital interactions. As a result, the orbitals are not static during the optimization process; they adapt to the specific problem.

A reconstruction functional for the 2RDM in terms of the ONs leads to the following electronic energy (GNOF): 
\begin{equation}
E_{el}=E^{intra}+E_{HF}^{inter}+E_{sta}^{inter}+E_{dyn}^{inter} \label{gnof}
\end{equation}

The intra-pair component is formed by summing the energies $E_{g}$ of electron pairs with opposite spins and the one-electron energies of unpaired electrons, specifically,
\begin{equation}
E^{intra}=\sum\limits _{g=1}^{\mathrm{N_{II}}/2}E_{g}+{\displaystyle \sum_{g=\mathrm{N_{II}}/2+1}^{\mathrm{N}_{\Omega}}}H_{gg}
\label{Eintra}
\end{equation}
\begin{equation}
E_{g} = 2 \sum\limits _{p\in\Omega_{g}}n_{p}H_{pp} + \sum\limits _{q,p\in\Omega_{g}} \Pi(n_q,n_p) L_{pq}
\end{equation}
\noindent where $H_{pp}$ are the diagonal one-electron matrix elements of the kinetic energy and external potential operators, whereas $L_{pq}=\left\langle pp|qq\right\rangle$ are the exchange-time-inversion integrals.\citep{Piris1999} The matrix elements $\Pi(n_q,n_p) = c(n_q)c(n_p)$, where $c(n_p)$ is defined by the square root of the ONs according to the following rule:
\begin{equation}
    c(n_p) = \left.
  \begin{cases}
    \phantom{+}\sqrt{n_p}, & p \leq \frac{\mathrm{N_{II}}}{2}\\
    -\sqrt{n_p}, & p > \frac{\mathrm{N_{II}}}{2} \\
  \end{cases}
  \right. \>\>, \quad p \in \Omega_{g} \in \Omega_{\mathrm{II}}
  \label{eq:PNOF5-roots}
\end{equation}
that is, the phase factor of $c_p$ is chosen to be $+1$ for the strongly occupied orbital of a given subspace $\Omega_g$, and $-1$ otherwise. The inter-subspace HF term is 
\begin{equation}
E_{HF}^{inter}=\sum\limits _{p,q=1}^{\mathrm{N}_{B}}\,'\,n_{q}n_{p}\left(2J_{pq}-K_{pq}\right)\label{ehf}
\end{equation}
where $J_{pq}=\left\langle pq|pq\right\rangle$ and $K_{pq}=\left\langle pq|qp\right\rangle $ are the Coulomb and exchange integrals, respectively. The prime in the summation indicates that only the inter-subspace terms are taking into account. The inter-subspace static component is written as 
\begin{equation}
E_{sta}^{inter} = -\left({\displaystyle \sum_{p=1}^{\mathrm{N}_{\Omega}} \sum_{q=\mathrm{N}_{\Omega}+1}^{\mathrm{N}_{B}} + \sum_{p=\mathrm{N}_{\Omega}+1}^{\mathrm{N}_{B}} \sum_{q=1}^{\mathrm{N}_{\Omega}} \> + \displaystyle \sum_{p,q = \mathrm{N}_{\Omega}+1}^{\mathrm{N}_{B}}} \right)' \Phi_{q}\Phi_{p}L_{pq} 
\nonumber
\end{equation}
\begin{equation}
- \: \dfrac{1}{2} \left({\displaystyle \sum\limits_{p=1}^{\mathrm{N_{II}}/2}\sum_{q=\mathrm{N_{II}}/2+1}^{\mathrm{N}_{\Omega}} \> + \left.{ \sum_{p=\mathrm{N_{II}}/2+1}^{\mathrm{N}_{\Omega}} \sum\limits _{q=1}^{\mathrm{N_{II}}/2}} \right)' \Phi_{q} \Phi_{p}L_{pq} {\displaystyle \:-\:\dfrac{1}{4}\sum_{p,q=\mathrm{N_{II}}/2+1}^{\mathrm{N}_{\Omega}}} K_{pq} }\right.
\label{esta}
\end{equation}
where $\Phi_{p}=\sqrt{n_{p}h_{p}}$ with the hole $h_{p}=1-n_{p}$. Finally, the inter-subspace dynamic energy is
\begin{equation}
E_{dyn}^{inter}=\sum\limits _{p,q=1}^{\mathrm{N}_{B}}\,'\,\left[n_{q}^{d}n_{p}^{d}+\;\Pi\left(n_{q}^{d},n_{p}^{d}\right)\right] \left(1-\delta_{q\Omega_{II}^{b}}\delta_{p\Omega_{II}^{b}}\right)L_{pq}
\label{edyn}
\end{equation}

In Eq. (\ref{edyn}), $\Omega_{II}^{b}$ denotes the subspace composed of orbitals below the level $\mathrm{N_{II}}/2$, and the dynamic part of the ON $n_{p}$ is defined as 
\begin{equation}
n_{p}^{d}=n_{p}\cdot e^{-\left(\dfrac{h_{g}}{h_{c}}\right)^{2}},\quad p\in\Omega_{g},\quad g=1,2,...,\frac{\mathrm{N_{II}}}{2} \label{dyn-on}
\end{equation}
with $h_{c}=0.02\sqrt{2}$. $n_{p}^{d}$ is in accordance with the Pulay\textquoteright s criterion that establishes an occupancy deviation of approximately 0.01 with respect to 1 or 0 for a NO to contribute to the dynamic correlation.

\begin{figure}[htbp]
\begin{centering}
\caption{Energetic profiles of a reactive trajectory for a translational energy of 2.45 eV and initial ro-vibrational state ($\nu=0,\mathrm{J}=0$) of H$_2$. \bigskip}
{\includegraphics[scale=0.3]{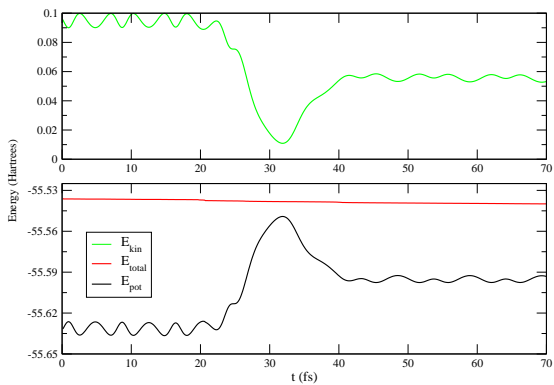}}
\label{eprof}
\end{centering}
\end{figure}

\begin{figure*}[htbp]
\centering
\caption{Time evolution of the two strongly occupied natural orbitals involved in the bond pattern change during the collision. \bigskip}
{\includegraphics[scale=0.4]{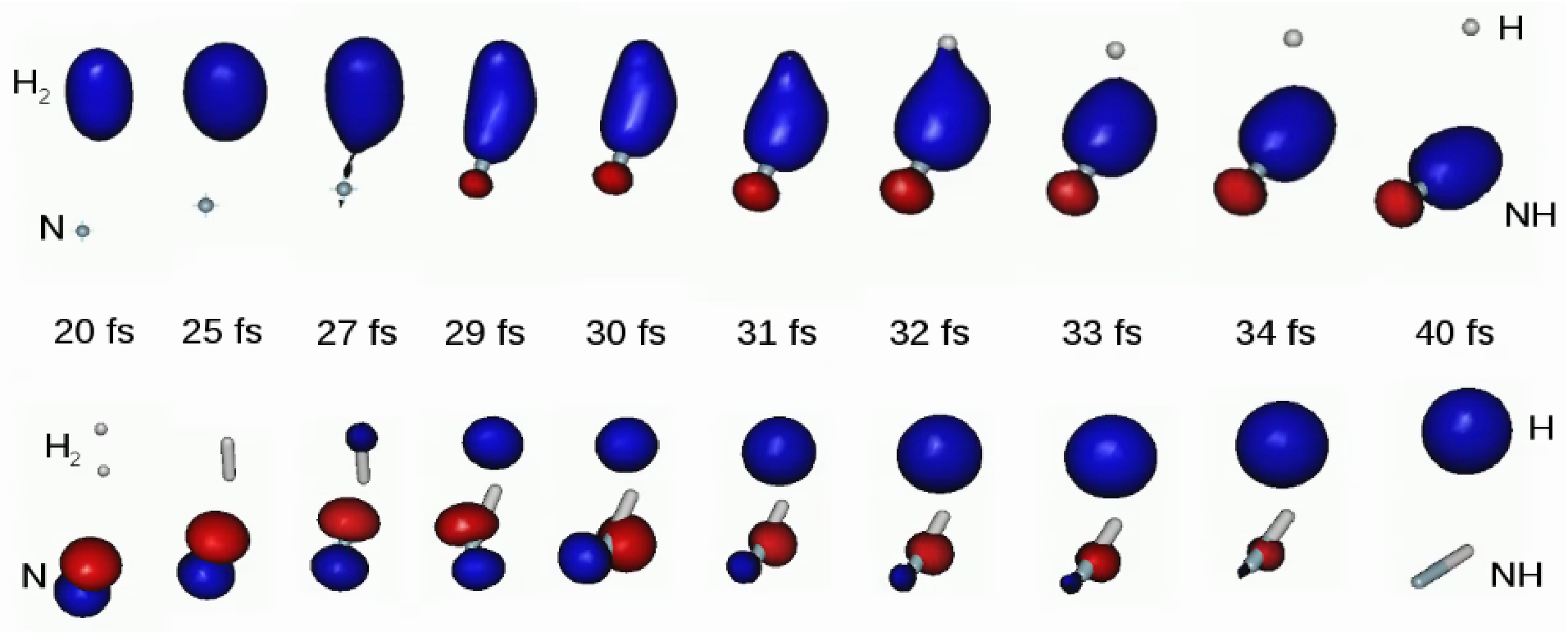}}
\label{orb-evol}
\end{figure*}

In the BO approximation, the total energy of the molecule can be cast as $E=E_{nuc}+E_{el}$ with $E_{nuc}=\sum_{A<B}[{Z_{A}Z_{B}}/{R_{AB}}]$. $Z_{A}$ represents the atomic number of nucleus $A$, and $R_{AB}$ is the distance between nuclei $A$ and $B$. Considering that all NOs $\phi_{i}$ are expanded in a fixed atomic basis set, $\phi_{i}=\sum_{\upsilon}{C}_{\upsilon i}\zeta_{\upsilon}$, the derivative of the total energy with respect to the coordinate $x$ of nucleus $A$ is given by \citep{Mitxelena2020c}
\begin{equation}
\begin{array}{c}
{\displaystyle \frac{dE}{dx_{A}}}={\displaystyle \frac{\partial E_{nuc}}{\partial x_{A}}} + {\displaystyle \sum_{\mu\upsilon}}\Gamma_{\mu\upsilon}\dfrac{\partial H_{\mu\upsilon}}{\partial x_{A}} + \\
{\displaystyle \sum_{\mu\eta\upsilon\delta}}D_{\mu\eta\upsilon\delta}\dfrac{\partial\left\langle \mu\eta|\upsilon\delta\right\rangle }{\partial x_{A}} - {\displaystyle \sum_{\mu\upsilon}}\lambda_{\mu\upsilon}\dfrac{\partial {S_{\mu\upsilon}}}{\partial x_{A}}.
\end{array}\label{NOF-grads}
\end{equation}
where $\Gamma_{\mu\upsilon}$ and $D_{\mu\eta\upsilon\delta}$ are the 1RDM and 2RDM, respectively, $S_{\mu\upsilon}=\left\langle \mu|\upsilon\right\rangle$ is the overlap matrix, and $\lambda_{\mu\upsilon}$ are the Lagrange multipliers obtained from RDMs, all in the atomic orbital representation. The first term of Eq. (\ref{NOF-grads}) is the derivative of the nuclear energy, the second represents the negative Hellmann-Feynman force, while the third contains the explicit derivatives of two-electron integrals. The last term, known as the density force, arises from the implicit dependence of $C_{\upsilon i}$ on geometry. The implicit dependence of ONs on geometry does not contribute to analytic gradients since $E_{el}$ is stationary with respect to variations in all of the ONs.\citep{Mitxelena2017}

All derivatives in Eq. (\ref{NOF-grads}) have an explicit dependence on the nuclear coordinate $x_{A}$, so the force acting on each nucleus A $(\textbf{F}_{A}=-\nabla_{A} E)$ can be obtained by a single static evaluation at each time step for the fixed nuclear positions at that instant. Consequently, we can calculate the trajectories of the nuclei according to the classical equations of motion, but taking into account the quantization of the reactants, a procedure known as quasiclassical trajectory (QCT) method. It is worth noting that the QCT method does not take the tunneling effect into account, so it can produce inaccurate results near the threshold energy.

NOF-based QCT calculations can be performed using the new molecular dynamics module implemented in DoNOF,\citep{Piris2021a} which allows the calculation of nuclear trajectories by determining ``on the fly'' the forces using NOF gradients (\ref{NOF-grads}). Beeman's algorithm\citep{Beeman1976} is used to numerically integrate Newton's equations of motion, whereas the initial conditions are obtained using a standard Monte Carlo sampling procedure.\citep{Truhlar1979}

The N($^4$S) + H$_2$($^1\Sigma$) $\rightarrow$ NH($^3\Sigma$) + H($^2$S) reaction using the cc-pVDZ basis set\citep{Dunning1989} has been taken as a validation test for the GNOF-based BO AIMD. This reaction is important in the decomposition of ingredients in solid propellants used for rockets,\citep{Prasad1997} so it has been the subject of several high-quality theoretical studies\citep{Zhang2000,Pascual2002,Poveda2005,Yu2013,Yu2014,Zhang2015} due to the experimental difficulty in preparing N atoms. These studies have shown that the reaction occurs via an abstraction mechanism dominated by the quartet GS, and presents a forward experimental barrier of 1.4 $\pm$ 0.3 eV.\citep{Koshi1990}

The initial separation between the nitrogen atom and the center-of-mass of H$_2$ was set at 6 \AA, and each trajectory was integrated until the separation between the final fragments was greater than 6~\AA. A time step of 0.1 fs was used, which yields to a conservation of the total energy with an average error of 0.004 eV. Typical profiles of the kinetic, potential and total energies for a reactive trajectory with a translational energy $\mathrm{E_T}$ = 2.45 eV and H$_2$ at the GS can be seen in Fig. \ref{eprof}, while in the supplementary material a movie of this trajectory can be found. From here we can conclude that the collision occurs mainly in the time range from 20 fs to 40 fs.

In the N($^4$S) + H$_2$($^1\Sigma$) reaction, six are the strongly occupied NOs, namely three with occupancy close to 2 and three singly occupied responsible for the quartet state. During reactive dynamics, the lowest energy orbitals correspond to the 1s and 2s atomic orbitals of N. These NOs start from the isolated atom and remain in the NH radical without significant changes. Similarly, two of the 2p atomic orbitals of N undergo small transformations during the collision and continue to maintain their character in the final NH, occupying directions perpendicular to the bonding direction. Consequently, two NOs are responsible for the change of the bond pattern during the collision, whose  temporal evolutions are represented in Fig. \ref{orb-evol} by specific snapshots. Thus, we have a first NO that begins as a $\sigma$ ``ss'' bonding orbital of the H$_2$ singlet with ON = 1.97 and transforms into the $\sigma$ ``sp'' bonding orbital of the NH triplet with ON = 1.96, while the other individually occupied NO transforms from a 2p atomic orbital of the N into the 1s atomic orbital of the H.

Reaction probabilities and integral cross sections (ICSs) were calculated for translational energies up to 5.0 eV, and five different initial ro-vibrational states ($\nu,\mathrm{J}$) of H$_2$ molecule, namely the GS (0,0), the first two excited vibrational states (1,0) and (2,0), and two excited rotational states (0,10) and (0,15). Reduced batches of 1000 trajectories were run to determine a first estimate of the maximum impact parameter values ($b_{\rm max}$). Finally, 5000 trajectories  were carried out using appropriate value of $b_{\rm max}$ for each translational energy and ro-vibrational state ($\nu,\mathrm{J}$) of the H$_2$ diatom, which provided the ICS values through the following equation:
\begin{equation}
\sigma\cong\pi b_{\rm max}^{2}\frac{N_r}{N_t},
\label{reaction-cross-section}
\end{equation}
\noindent where $N_r$ is the number of reactive trajectories, $N_{t}$ the total number of trajectories for which the impact parameter satisfies $b\le b_{\rm max}$, and $P_r=N_r/N_t$ is the total reaction probability. The number of propagated trajectories ensures a Monte Carlo statistical error of less than $5\%$ for the ICS. Additionally, the final energy distribution of the product was obtained after a standard semi-classical determination of the vibrational, and rotational energies\cite{Truhlar,miller1974classical,gw1997}.

\begin{figure}[htbp]
\begin{centering}
\caption{Maximum impact parameter ($b_{\rm max}$) and integral cross section ($\sigma$) as a function of translational energy ($\mathrm{E_T}$) for different H$_{2}$ ro-vibrational states ($\nu,\mathrm{J}$). Dashed lines correspond to values reported in Ref.\citep{Yu2014} \bigskip \bigskip } 
{\includegraphics[scale=0.36]{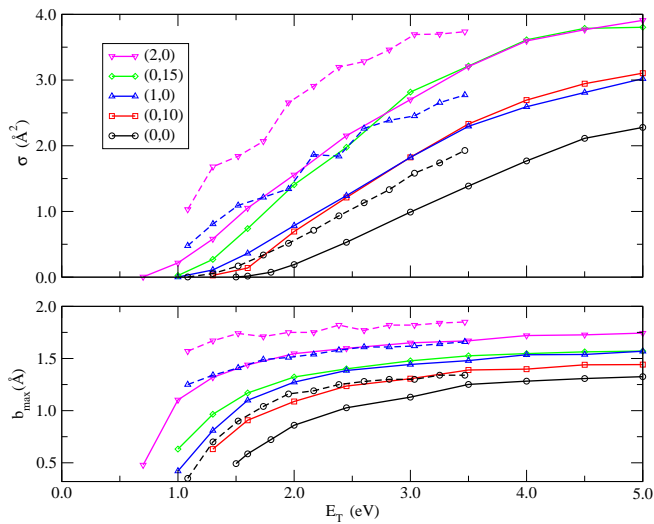}}
\label{bmax-ics}
\end{centering}
\end{figure}

Fig. \ref{bmax-ics} presents $b_{\rm max}$ and $\sigma$ as a function of $\mathrm{E_T}$ for the five ro-vibrational states studied. For comparison, values reported in Ref.\citep{Yu2014} based on the highly accurate DMBE PES\citep{Poveda2005} calculated at the MRCI-FVCAS/aug-cc-pVQZ level of theory are included. The $b_{\rm max}$ values were found to vary from 0.42~\AA~to 1.74~\AA~on increasing the collision energy. For the H$_2$ GS, the figure shows that the ICS has a threshold energy of approximately 1.5 eV and gradually decreases with increasing $\nu$ and J to 0.7 and 0.97 eV, respectively, in perfect agreement with the forward barrier observed for the reaction. Above the threshold energy, the ICSs increase almost linearly up to around 4 eV, where a much slower growth begins. For the ICSs corresponding to initial rotationally excited H$_2$ (J = 10,15), we observe that a decay then begins. Notice that the initial rotational excitation of H$_2$ molecule cause a more pronounced linear increase in ICSs at lower collision energies, and that the increase is significant in reactivity with increasing reagent vibrational excitation, as expected in this endoergic triatomic reaction with a late barrier. The shape of the ICS curves presented in Fig. \ref{bmax-ics} is similar to what was observed in previous studies, however our ICS values are lower. This ICS underestimation is understandable considering the modest basis set used in this work that lead to lower values of $b_{\rm max}$. 

\begin{figure}[htbp]
\begin{centering}
\caption{Rotational-state distribution of product NH for the reaction N + H$_2$($\nu=0,\mathrm{J}=0$) $\rightarrow$ NH ($\nu',\mathrm{J}'$) + H. \bigskip \bigskip \bigskip } 
{\includegraphics[scale=0.36]{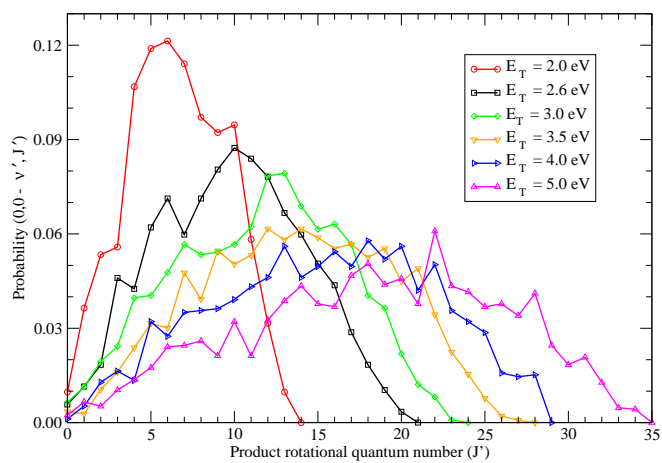}}
\label{jdist}
\end{centering}
\end{figure}

Regarding the final ro-vibrational energy distribution of the product when the H$_2$ molecule is initially in its ground state ($\nu=0,\mathrm{J}=0$) our simulations show that below 3 eV, only the vibrational ground state ($v'=0$) and the first vibrational level ($v'=1$) are accessible to the NH molecule after its formation. At the highest collision energy studied in this work (5 eV), the highest vibrational level populated was $v'=5$. These results indicate a poor transfer between the translational and vibrational energy modes during the collision. However, an efficient rotational excitation of the NH molecule for the studied range of collision energies was observed. For a collision energy of 5 eV, 35 rotational levels are accessible, while for 1.5 eV, the six first rotational levels were populated. We show in Fig. \ref{jdist} the calculated rotational-state distributions of the product for the reaction N + H$_2$($\nu=0,\mathrm{J}=0$) $\rightarrow$ NH($v'$,J$'$) + H for six collision energies. The figure clearly indicates that the rotational excitation increases with increasing $\mathrm{E_T}$. The higher the collision energy of the reactants, the higher the rotational energies of the product obtained. This behaviour is in prefect agreement with previous theoretical works\cite{Pascual2002,Yu2013} on the studied reaction. The results presented in figures \ref{bmax-ics} and \ref{jdist} emphasize the sensitivity of our method to qualitatively and quantitatively fully captures the physics involve in the dynamics of complex reactions.

In summary, we have shown that GNOF-AIMD is a method of choice to investigate the evolution of complex electronic problems, particularly reactive collisions. According to the existence theorems\citep{Gilbert1975, Levy1979, Valone1980} of the 1RDM functional, there is a one-to-one mapping between the ground-state 1RDM and the ground-state N-particle density matrix, so by observing the real-time evolution of the NOs along with their ONs, we are seeing in real time the evolution of the solution to the electronic problem dynamically. The unique NO representation is especially useful for viewing the real-time evolution of changes in bond patterns. Indeed, NOs vary along trajectories calculated on the fly, adapting at each time step to the most favorable interactions of the corresponding nuclei configuration. GNOF-AIMD also allows to study dynamics with any transformation in the spins and the number of electrons of the component subsystems, conserving the total spin of the entire system. Therefore, GNOF-AIMD opens a promising field of research: AIMD based on natural orbital functionals.

\bigskip
\textbf{Acknowledgements}: Support comes from Grant PID 2021-126714NB-I00 funded by MCIN/AEI/10.13039/ 501100011033 and the Eusko Jaurlaritza (Ref.: IT1584-22). The authors thank for technical and human support provided by IZO-SGI SGIker of UPV/EHU and DIPC.



%

\end{document}